\documentstyle[12pt]{article}
\begin{document}
\mbox{ }
\rightline{UCT-TP-221/94}
\rightline{November 1994}

\begin{center}

{\Large \bf Phenomenological analysis of a dimension-two\\[.5cm]
 operator in QCD and its impact on $\alpha_{s}(M_{\tau})$}\\[1.5cm]

{\large \bf C. A. Dominguez}\footnote{John Simon Guggenheim Fellow
1994-1995}\\[.5cm]

Institute of Theoretical Physics and Astrophysics, University of Cape
Town, Rondebosch 7700, South Africa\\
\end{center}

\vspace{1.5cm}

\begin{abstract}
\noindent
Fits to the ARGUS data on hadronic decays of the tau-lepton, which
determine the vector and axial-vector spectral functions, are used
in order to determine the size of a dimension $d=2$ term in the
Operator Product Expansion. Constraints from the first Weinberg sum
rule (in the chiral limit) are enforced in order to reduce the uncertainty
of this determination. Results for  the $d=2$ operator
are consistent with a quadratic dependence on $\Lambda_{QCD}$. The impact
of this term on the extraction of $\alpha_{s}(M_{\tau})$ is assessed.
\end{abstract}

\setlength{\baselineskip}{1.25\baselineskip}
\newpage
A technique based on analyticity and the Operator Product Expansion (OPE)
of current correlators was proposed some time ago \cite{P1} to extract
the strong coupling constant from data on $e^{+} e^{-}$ annihilation,
and on the total hadronic width of the tau-lepton. Recent results \cite{P2}
suggest this method to be unrivaled in precision. However, this high
precision has been questioned recently \cite{ALT} on the grounds that it
relies on the assumption that no operators of dimension $d=2$
(other than well known quark-mass insertion terms) enter the
OPE of the two-point functions involving the vector and the axial-vector
currents. While it is not possible to construct vacuum condensates of $d=2$
with the quark and gluon fields of the QCD Lagrangian, one cannot e.g. rule
out a-priori a term of the form $C_{2} \propto \Lambda_{QCD}^{2}$. In fact,
such terms have been suggested as resulting from large order perturbation
theory \cite{REN}. Given the importance of this issue it becomes imperative
to analyze the existing experimental data in order to decide if the
presence of such a $d=2$ term can be established. An important step
in this direction has been taken in \cite{NAR} by determining the size
of this term from experimental data on  $e^{+} e^{-}$ annihilation
into hadrons, together with a zeroth-moment Finite Energy Sum Rule (FESR),
and a Laplace transform QCD sum rule.\\

In this note I analyze this problem using the ARGUS data \cite{AR} on hadronic
decays of the tau-lepton, which determine the hadronic spectral functions
in the (non-strange) vector and axial-vector channels. These data have
been used in the past \cite{DS1} to extract the values of the vacuum
condensates of dimension $d=$ 4, 6 and 8 (assuming no term of $d=2$ other
than the one from the quark-mass insertion), as well as to check the
saturation of the  Weinberg sum rules in the chiral limit \cite{DS2}. I shall
perform the analysis by means of the zeroth-moment FESR, which has the
advantage of decoupling the vacuum condensates of $d > 2$. In contrast,
Laplace transform sum rules involve condensates of all dimensions whose
values are affected by very large uncertainties. In addition, I shall
consider the constraint imposed by the first Weinberg sum rule on the
integrals of the vector and axial-vector spectral functions. This procedure
effectively reduces the uncertainties induced by the experimental errors.
I shall proceed without prejudice on the uncertainty in
$\alpha_{s}(M_{\tau})$, and determine $C_{2}$ separately for two different
values of $\Lambda_{QCD}$, viz. $\Lambda_{QCD} = 100$ MeV,
and $\Lambda_{QCD} = 300$ MeV. This will allow to establish
a possible functional dependence of $C_{2}$ on $\Lambda_{QCD}$.
Use of the tau-lepton decay data offers a comparative advantage over
$e^{+} e^{-}$ annihilation in  that $C_{2}$ determined from
the vector channel can be confronted with the (independent) result from
the axial-vector channel. Since chirality considerations require both values
to be the same, one is offered a consistency check and a means of
reducing the uncertainty on $C_{2}$.\\
The relevant two-point functions needed for this analysis are
\begin{eqnarray}
\Pi_{\mu \nu}^{V} \; (q) &=& i \int d^{4}x e^{i q x} <0| \; T(V_{\mu}(x) \;
V_{\nu}^{\dagger}(0))|0>\\[.3cm] \nonumber
  &=& - (g_{\mu \nu} q^{2} - q_{\mu} q_{\nu}) \Pi_{V} (q^{2}) \,\,,
\end{eqnarray}
\begin{eqnarray}
\Pi_{\mu \nu}^{A} \; (q) &=& i \; \int \; d^{4}x \; e^{i q x} \; <0| \;
 T(A_{\mu}(x) \; A_{\nu}^{\dagger}(0))|0>\\[.3cm] \nonumber
  &=& - g_{\mu \nu} \; \tilde{\Pi}_{A} \; (q^{2}) + q_{\mu} q_{\nu} \;
\Pi_{A} (q^{2}) \,\,,
\end{eqnarray}
where $V_{\mu} = (\bar{u} \gamma_{\mu} u - \bar{d} \gamma_{\mu} d)/2$,
and  $A_{\mu} = (\bar{u} \gamma_{\mu} \gamma_{5} u - \bar{d} \gamma_{\mu}
\gamma_{5} d)/2$. Considering these (charge neutral) currents implies
the normalization: $Im \Pi_{V} = Im \Pi_{A} = 1/8\pi$, at one loop order
in perturbative QCD. This choice of normalization facilitates comparison with
the $e^{+} e^{-}$ channel. The lowest moment FESR, which projects out
a dimension $d=2$ term in the OPE, is of the form
\begin{eqnarray}
I_{0} (s_{0})_{V,A} \equiv \frac{8 \pi^{2}}{s_{0}} \;
\int_{0}^{s_{0}} \; \frac{1}{\pi} \; \mbox{Im} \; \Pi_{V,A} (s) \; ds =
F_{2} (s_{0}) + \frac{C_{2}^{V,A}}{s_{0}} \,\,,
\end{eqnarray}
where the radiative corrections to 4-loop order can be written as
\begin{eqnarray*}
F_{2} (s_{0}) = 1 + \frac{\alpha_{s}^{(1)}(s_{0})}{\pi} +
\left( \frac{\alpha_{s}^{(1)} (s_{0})}{\pi} \right)^{2} \;
\left( F_{3} - \frac{\beta_{2}}{\beta_{1}} \; \ln \; L -
\frac{\beta_{1}}{2} \right)
\end{eqnarray*}
\begin{eqnarray*}
+ \left( \frac{\alpha_{s}^{(1)} (s_{0})}{\pi} \right)^{3} \;
\left[ \frac{\beta_{2}^{2}}{\beta_{1}^{2}} \; (\ln^{2} \; L - \ln \;
L -1) + \frac{\beta_{3}}{\beta_{1}} \right.
\end{eqnarray*}
\begin{eqnarray}
\left. - 2 \; \left( F_{3} - \frac{\beta_{1}}{2} \right)
\; \frac{\beta_{2}}{\beta_{1}} \; \ln \; L + F_{4} - F_{3}
\beta_{1} - \frac{\beta_{2}}{2} + \frac{\beta_{1}^{2}}{2}  \right] \; ,
\end{eqnarray}
with
\begin{eqnarray}
\frac{\alpha_{s}^{(1)}(s_{0})}{\pi} \equiv
\frac{- 2}{\beta_{1} L} \,\, ,
\end{eqnarray}
where $L \equiv ln (s_{0}/\Lambda_{QCD}^{2})$,
and for two flavours: $\beta_{1} = - 29/6$, $\beta_{2} = - 115/12$,
$\beta_{3} = - 48241/1728$, $F_{3} = 1.756$, $F_{4} = - 9.057$.
In writing Eq. (4) use has been made  of the result \cite{ALPHA}
\begin{eqnarray}
\frac{\alpha_{s}^{(3)}}{\pi} = \frac{\alpha_{s}^{(1)}}{\pi} +
\left( \frac{\alpha_{s}^{(1)}}{\pi} \right)^{2}
\left(- \frac{\beta_{2}}{\beta_{1}} \; \ln \; L \right) +
\left( \frac{\alpha_{s}^{(1)}}{\pi} \right)^{3}
\left[ \frac{\beta_{2}^{2}}{\beta_{1}^{2}}
\; (\ln^{2} L - \ln \; L - 1) +
\frac{\beta_{3}}{\beta_{1}}  \right] \,\,.
\end{eqnarray}
Notice that the convention for the sign of the $d=2$ term in Eq.(3)
is opposite to the one used in \cite{NAR}.
At dimension $d=2$ there is a well known mass insertion contribution to
$C_{2}$, e.g. in the vector channel it is given by
\begin{equation}
C_{2 m} = - 3 \frac{(\hat{m}_{u}^{2} + \hat{m}_{d}^{2})}
{(\frac{1}{2} \ln s_{0}/\Lambda_{QCD}^{2})^{-4/\beta_{1}}}
\end{equation}
Using standard values for the current up and down-quark masses \cite{DR}
this term is negligible. Hence, in the sequel the chiral limit will be
used throughout.\\

The analysis of the ARGUS data \cite{AR} to extract the vector and
axial-vector spectral functions entering Eq.(3) has been discussed in
\cite{DS1}. Using these fits to the data and performing the integrations
leads to the results shown in Fig.1 for $I_{0}(s_{0})_{V}$, and in Fig.2
for $I_{0}(s_{0})_{A}$ (solid curves). The dashed lines correspond to
$F_{2}(s_{0})$, Eq.(4), for $\Lambda_{QCD} = 100$ MeV (lower line) and
$\Lambda_{QCD} = 300$ MeV (upper line). The error bar is an estimate of
the propagation of the experimental errors in the dispersive integrals.
The agreement between the results of the fits and perturbative QCD is
rather good in the interval $1.5 \mbox{GeV}^{2} < s_{0} < 2.5 \mbox{GeV}^{2}$,
although there is some room left to acommodate a non-zero value of $C_{2}$.
It is possible to reduce effectively the impact of the experimental
uncertainties by considering the first Weinberg sum rule in the chiral limit
\begin{eqnarray}
W_{1} (s_{0}) \equiv \int_{0}^{s_{0}} \;
\left[ \frac{1}{\pi} \mbox{Im} \; \Pi_{V}(s) - \frac{1}{\pi} \; \mbox{Im} \;
\Pi_{A} (s) \right] \; ds \,\,,
\end{eqnarray}
where the pion-pole pole contribution, equal to $f_{\pi}^{2}$
($f_{\pi} = 93.2$ MeV), is already included in the axial-vector spectral
function. Using the fits to the data in Eq.(8) gives the result shown
in Fig.3, which indicates a very good saturation of the sum rule.
Notice that while strictly speaking only $W_{1}(\infty) \equiv 0$,
for $s_{0} > 2.5 \mbox{GeV}^{2}$ both spectral functions are well
approximated by their (identical) perturbative QCD expressions, as
evidenced by the results of the fits (for more details see
\cite{DS1}-\cite{DS2}). Hence, this sume rule becomes effectively a FESR.
{}From $C_{2}^{V} = C_{2}^{A}$, this term does not enter Eq.(8). Hence, the
first Weinberg sum rule provides an independent constraint on the areas
under the vector and axial-vector spectral functions.\\
After confronting the left- and right-hand sides of Eq.(3) one obtains
for $C_{2}$ the results shown in Fig.4. The upper set of curves corresponds
to $\Lambda_{QCD} = 100$ MeV, and the lower set to
$\Lambda_{QCD} = 300$ MeV. Curves (a) and (b) stand for $C_{2}$ obtained
from the axial-vector and the vector channel, respectively. Using
$C_{2}^{V} = C_{2}^{A}$, independent of $s_{0}$, determines an
overlap region (for each value of $\Lambda_{QCD}$) from which the
following values are obtained
\begin{eqnarray}
C_{2} (\Lambda_{QCD} = 100 \mbox{MeV}) &=& - (0.028 \pm 0.012) \mbox{GeV}^{2}
\nonumber
\\
C_{2} (\Lambda_{QCD} = 300 \mbox{MeV}) &=& - (0.200 \pm 0.006) \mbox{GeV}^{2}.
\end{eqnarray}
Taking the ratio of the above results gives
\begin{equation}
\frac{C_{2} (\Lambda_{QCD} = 300 \mbox{MeV})}
{C_{2} (\Lambda_{QCD} = 100 \mbox{MeV})} = 7 \pm 3 \, ,
\end{equation}
which is consistent with a ratio of 9, or $C_{2} \propto \Lambda_{QCD}^{2}$,
as if it would originate from renormalons \cite{REN}.\\
With the sign convention used here, the result found in \cite{NAR} using
the $e^{+} e^{-}$ data in the FESR is $C_{2} = + 0.02 \pm 0.12
\mbox{GeV}^{2}$, for an input value $\alpha_{s}(M_{\tau}) = 0.32 \pm
0.04$. If a more conservative error in $\alpha_{s}(M_{\tau})$ had been
used, e.g. some 50\% bigger, then the value of $C_{2}$ from that analysis
would have been consistent with the one obtained here, except for the
sign. It is possible to understand the sign difference from the following
observation. It is known from the $e^{+} e^{-}$ analysis of \cite{BERT}
that $I_{0}(s_{0})_{V}$ approaches the perturbative QCD result
$F_{2}(s_{0})$ from {\bf above}, in contrast with the case of the
tau-lepton analysis, where it approaches it from {\bf below}. Inside the
duality region, though, the areas under the vector spectral functions
from the two analyses are consistent with each other. If one were to use
$Im \Pi_{V}$ extracted from $e^{+} e^{-}$ data, and $Im \Pi_{A}$
from tau-lepton decays, then the first Weinberg sum rule would be
saturated from above, rather than from below, as in Fig. 3. Hence,
considering both analyses together makes the sign of $C_{2}$ indeterminate.\\
The current value of $\alpha_{s}(M_{\tau})$ from the method of \cite{P1}
is \cite{P3}
\begin{equation}
\alpha_{s}(M_{\tau}) = 0.36 \pm 0.03 \,\,.
\end{equation}
This is obtained from the ratio of the hadronic to the leptonic widths
of the $\tau$-lepton, $R_{\tau}$, which can be written as
\begin{equation}
R_{\tau} = 3 (1 + \delta^{0} + \delta^{2}_{m} + \delta^{6} +
\delta^{8} + ...) \,\,,
\end{equation}
where $\delta^{0}$ contains the perturbative corrections, $\delta^{2}_{m}$
is the quark-mass insertion at $d=2$, $\delta^{6}$ stands for the
$d=6$ vacuum condensate contribution in the OPE, and so on. Corrections
from the electro-weak sector can also be added to Eq.(12).
The uncertainty in Eq.(11) includes, in addition to the experimental
error on $R_{\tau}$, perturbative and non-perturbative
uncertainties (except for a $d=2$ term different from $C_{2m}$). At a
scale $s_{0} = M_{\tau}^{2}$, which is not too far away from the duality
region found in this analysis, the results given in Eq.(9) imply a
correction of up to 6 \% in the QCD evaluation of the $\tau$ hadronic
width. This is not particularly negligible. While this correction
is still a factor of 3 smaller than the perturbative
QCD contribution (for $\delta^{0} \simeq 0.20$),
it may still have an impact on the final uncertainty in
$\alpha_{s}(M_{\tau})$. For instance, for $\Lambda_{QCD} = 300 \mbox{MeV}$
the error in $\alpha_{s}(M_{\tau})$ could go up to $\pm 0.06$, which is
a non-negligible 100 \% effect. A better theoretical understanding of
the origin of the $d=2$ operator would clearly help in placing further
constraints on its numerical contribution to Eq.(12).
\subsection*{Acknowledgements}
The author wishes to thank Guido Altarelli for a discussion which
triggered this research.
This work was supported in part by the John Simon Guggenheim Memorial
Foundation (USA) and the Foundation for Research Development (ZA).

\subsection*{Figure Captions}
\begin{description}
\begin{itemize}
\item[Figure 1:]
The left-hand side of Eq.(3) in the vector channel (solid curve), and
the perturbative QCD term $F_{2}(s_{0})$ for $\Lambda_{QCD} = 300 \mbox{MeV}$
(upper broken line), and $\Lambda_{QCD} = 100 \mbox{MeV}$ (lower broken
line).
\item[Figure 2:]
The left-hand side of Eq.(3) in the axial-vector channel (solid curve), and
the perturbative QCD term $F_{2}(s_{0})$ for $\Lambda_{QCD} = 300 \mbox{MeV}$
(upper broken line), and $\Lambda_{QCD} = 100 \mbox{MeV}$ (lower broken
line).
\item[Figure 3:]
The Weinberg sum rule in the chiral limit, Eq.(8).
\item[Figure 4:]
Curves (a) and (b) stand for $C_{2}$ obtained
from the axial-vector and the vector channel, respectively.
The upper set of curves corresponds
to $\Lambda_{QCD} = 100$ MeV, and the lower set to
$\Lambda_{QCD} = 300$ MeV.
\end{itemize}
\end{description}
\end{document}